\documentclass[floats,floatfix,twocolumn,superscriptaddress,nofootinbib,showpacs]{revtex4-1}

\pdfoutput=1
\usepackage{amsfonts,amssymb,amsmath}
\usepackage[usenames,dvipsnames]{xcolor}
\usepackage[unicode,colorlinks=true,linkcolor=blue,urlcolor=blue,citecolor=gray]{hyperref}
\usepackage[all]{hypcap}
\usepackage{graphicx}
\usepackage{xspace}
\usepackage{tikz}
\usetikzlibrary{shapes,arrows}
\usetikzlibrary{matrix}
\usetikzlibrary{shapes.multipart}
\usetikzlibrary{er,positioning}
\usepackage[normalem]{ulem} 
\usepackage{url}
\usepackage{color,soul}
\usepackage[caption=false]{subfig}
\usepackage[T1]{fontenc}
\usepackage[utf8]{inputenc}
\usepackage{microtype}

\newcommand{\Flatiron}{\affiliation{Center for Computational Astrophysics, Flatiron Institute, 162 5th Ave, New York, NY 10010}}

\newcommand{\eps}{\ensuremath{\varepsilon}}

\newcommand{\pd}{\partial}

\newcommand{\nn}{\nonumber}

\newcommand{\txt}[1]{{\textrm{\tiny{#1}}}}
\newcommand{\mpl}{\ensuremath{m_\txt{pl}}}

\newcommand{\dual}{\,{}^*\!}  
\newcommand{\agb}{\alpha_\mathrm{GB}}

\begin{document}

\title{Stability of rotating black holes in Einstein dilaton Gauss-Bonnet gravity}

\author{Maria Okounkova}
\email{mokounkova@flatironinstitute.org}
\Flatiron

\date{\today}

\begin{abstract}
In order to perform model-dependent tests of general relativity with gravitational wave observations, we must have access to numerical relativity binary black hole waveforms in theories beyond general relativity (GR). In this study, we focus on order-reduced Einstein dilaton Gauss-Bonnet gravity (EDGB), a higher curvature beyond-GR theory with motivations in string theory. The stability of single, rotating black holes in EDGB is unknown, but is a necessary condition for being able to simulate binary black hole systems (especially the early-inspiral and late ringdown stages) in EDGB. We thus investigate the stability of rotating black holes in order-reduced EDGB. We evolve the leading-order EDGB scalar field and EDGB spacetime metric deformation on a rotating black hole background, for a variety of spins. We find that the EDGB metric deformation exhibits linear growth, but that this level of growth exponentially converges to zero with numerical resolution. Thus, we conclude that rotating black holes in EDGB are numerically stable to leading-order, thus satisfying our necessary condition for performing binary black hole simulations in EDGB. 
\end{abstract}

\maketitle

\section{Introduction}

At some length scale, Einstein's theory of general relativity (GR) must break down and be reconciled with quantum mechanics in a beyond-GR theory of gravity. Binary black hole systems probe the strong-field, non-linear regime of gravity, and thus gravitational wave signals from these systems could potentially contain signatures of a beyond-GR theory. The Laser Interferometry Gravitational Wave Observatory (LIGO) presently performs model-independent and parametrized tests of general relativity~\cite{TheLIGOScientific:2016src, LIGOScientific:2019fpa}, but one additional avenue of looking for deviations from general relativity is to perform \textit{model-dependent} tests. There has thus been an effort in recent years in performing numerical relativity simulations of binary black holes in beyond-GR theories~\cite{Healy:2011ef, Okounkova:2017yby, Witek:2018dmd, MashaHeadOn}.

In this study, we focus on numerical relativity in Einstein dilaton Gauss-Bonnet (EDGB) gravity, an effective field theory that modifies the Einstein-Hilbert action of GR through the inclusion of a scalar field coupled to terms quadratic in curvature that are meant to encompass underlying quantum gravity effects. In particular, EDGB is motivated by string theory~\cite{1987NuPhB.291...41G, Moura:2006pz, Berti:2015itd}. The well-posedness of the initial value problem in full, non-linear EDGB gravity is unknown~\cite{Delsate:2014hba}. We can, however work in an \textit{order-reduction scheme}, in which we perturb the EDGB scalar field and metric about a GR background. At each order, the equations of motion are well-posed (cf.~\cite{Witek:2018dmd}). 

In~\cite{Witek:2018dmd}, Witek et al. evolved the leading-order EDGB scalar field on a binary black hole background. In this study, we work one order higher, considering the leading-order EDGB deformation to the spacetime metric. In particular, we consider the behavior of this metric deformation on single black hole spacetimes. 

While our ultimate goal is to produce leading-order EDGB corrections to a binary black hole gravitational waveform (as we have done in dynamical Chern-Simons (dCS) gravity, another quadratic theory of gravity~\cite{MashaHeadOn}), we must first consider the question of black hole stability in EDGB. If the EDGB metric deformation around an isolated, rotating black hole is not stable, and exhibits pathological growth in time, then we know that we will not be able to simulate black hole binaries in order-reduced EDGB. The metric deformation around each black hole would grow during inspiral, and on the remnant black hole after merger, thus spoiling the evolution.

The analytical stability of black holes in EDGB is unknown. An exact static black hole solution in EDGB~\cite{Kanti:1995vq} is known, as are solutions in the slow-rotation limit~\cite{Pani:2009wy,Ayzenberg:2014aka}, rapid-rotation limit~\cite{Kleihaus:2011tg}, and full theory~\cite{Kleihaus:2015aje, Cunha:2016wzk}. The BH stability of radial and axial perturbations of static BHs has been studied~\cite{Kanti:1997br, Pani:2009wy}. Full stability of rotating black holes in EDGB is an open problem~\cite{Berti:2015itd}.

In this paper, we repeat the analysis of~\cite{MashaEvPaper}, in which we showed leading-order stability of rotating black holes in dCS, for EDGB. We evolve the leading-order EDGB metric deformation on a rotating black hole background using the formalism of~\cite{MashaEvPaper} for a variety of spins, and find that it is numerically stable. 

\subsection{Conventions}

We set $G = c = 1$ throughout. Quantities are given in terms of  units of $M$, the ADM mass of the Kerr black hole background. Latin letters in the beginning of the alphabet $\{a, b,  c, d \ldots \}$ denote 4-dimensional spacetime indices, and $g_{ab}$ refers to the spacetime metric with covariant derivative $\nabla_c$.

\section{Numerical setup}

\subsection{Equations of motion}

The overall form of the EDGB action that we will use in this paper, chosen to be consistent with Witek et al.~\cite{Witek:2018dmd}, is

\begin{align}
    S \equiv \int \frac{\mpl^2}{2} d^4 x \sqrt{-g} \left[R - \frac{1}{2} (\pd \vartheta)^2 + 2 \alpha_\mathrm{GB} f(\vartheta) \mathcal{R}_\mathrm{GB} \right]\,,
\end{align}
where the first term is the familiar Einstein-Hilbert action of GR (where $R$ is the 4-dimensional Ricci scalar), $\vartheta$ is the EDGB scalar field, and $\alpha_\mathrm{GB}$ is the EDGB coupling parameter with dimensions of length squared. Finally, $\mathcal{R}_\mathrm{GB}$ is the EDGB scalar, of the form 
\begin{align}
\label{eq:RGBDefinition}
    \mathcal{R}_\mathrm{GB} = R^{abcd}R_{abcd} - 4 R^{ab} R_{ab} + R^2\,.
\end{align}

In order to ensure well-posedness (cf.~\cite{Delsate:2014hba}), we perturb the spacetime metric and EDGB scalar field about an arbitrary GR background as
\begin{align}
    g_{ab} &= g_{ab}^{(0)} + \sum_{n = 1}^{\infty} \eps^n g_{ab}^{(n)}\,, \\
    \vartheta &= \sum_{n = 0}^\infty \eps^n \vartheta^{(n)}\,,
\end{align}
where $\eps$ is an order-counting parameter that counts powers of $\agb$, and ${(0)}$ corresponds to the GR solution. The quantity $\vartheta^{(0)}$ can be freely set to zero, and thus $g_{ab}^{(1)}$ vanishes (cf.~\cite{Witek:2018dmd}). The leading-order dynamics thus occurs in $\vartheta^{(1)}$ and $g_{ab}^{(2)}$, which are precisely the fields we evolve in this study.

The equation of motion for $\vartheta^{(1)}$, the leading (first) order EDGB scalar field, takes the form (cf.~\cite{Witek:2018dmd} for a full derivation),
\begin{align}
    \square^{(0)} \vartheta^{(1)} &= - M^2 \mathcal{R}^{(0)}_\mathrm{GB}\,, \\
    \label{eq:RGB0Definition}
     \mathcal{R}^{(0)}_\mathrm{GB} &\equiv R^{(0)} {}^{abcd}R^{(0)}_{abcd} - 4 R^{(0)} {}^{ab} R^{(0)}_{ab} + R^{(0)} {}^2\,,
\end{align}
where the superscript ${(0)}$ refers to quantities computed from the GR background. Here, the leading-order correction to $f(\vartheta)$ has been set to $\frac{1}{8}$, corresponding to EDGB gravity~\cite{Witek:2018dmd}. 

Meanwhile, the equation for $g_{ab}^{(2)}$, the leading (second) order EDGB deformation to the spacetime metric takes the form (cf.~\cite{Witek:2018dmd}),
\begin{align}
\label{eq:DeltapsiEOM}
    G_{ab}^{(0)} [g_{ab}^{(2)}] = - 8 M^2 \mathcal{G}_{ab}^{(0)}[\vartheta^{(1)}] + T_{ab}[\vartheta^{(1)}]\,.
\end{align}

In the above equations, $T_{ab}[\vartheta^{(1)}]$ is the standard Klein-Gordon stress energy tensor associated with $\vartheta^{(1)}$, of the form\footnote{Note that our definition of $T_{ab}[\vartheta^{(1)}]$ in Eq.~\eqref{eq:Tab1} differs from Eq. 15 in~\cite{Witek:2018dmd} by a factor of $2$, and hence the $T_{ab}[\vartheta^{(1)}]$ term in Eq.~\eqref{eq:DeltapsiEOM} differs by a factor of $2$. We have chosen this convention to be in line with the canonical form of the Klein-Gordon stress-energy tensor.}
\begin{align}
\label{eq:Tab1}
    T_{ab}[\vartheta^{(1)}] = \nabla^{(0)}_a \vartheta^{(1)} \nabla^{(0)}_b \vartheta^{(1)} - \frac{1}{2} g_{ab}^{(0)} \nabla^{(0)}_c \vartheta^{(1)} \nabla^{(0)}{}^c \vartheta^{(1)}\,,
\end{align}
and 
\begin{align}
\label{eq:G0Definition}
    \mathcal{G}_{ab}^{(0)}[\vartheta^{(1)}] = 2 \epsilon^{edfg} g_{c(a}^{(0)} g_{b)d}^{(0)} \nabla_h^{(0)} \left[\frac{1}{8} \dual R^{(0)}{}^{ch} {}_{fg} \nabla^{(0)}_e \vartheta^{(1)} \right]\,,
\end{align}
where $\dual R^{ab} {}_{cd} = \epsilon^{abef} R_{efcd}^{(0)}$ and $\epsilon^{abcd}$ is the Levi-Citiva pseudo-tensor, with $\epsilon^{abcd} = -[abcd]/\sqrt{-g^{(0)}}$, where $[abcd]$ is the alternating symbol.  

Note that on a vacuum background spacetime (as in the Kerr BH spacetime in our case), terms in $\mathcal{R}_\mathrm{GB}^{(0)}$ (cf. Eq.~\eqref{eq:RGB0Definition}) and $\mathcal{G}^{(0)}_{ab}[\vartheta^{(1)}]$ (cf. Eq.~\eqref{eq:G0Definition}) vanish to give the simplified equations of motion 
\begin{align}
    \square^{(0)} \vartheta^{(1)} &= - M^2 R^{(0)} {}^{abcd}R^{(0)}_{abcd} \\
    \nn G_{ab}^{(0)} [g_{ab}^{(2)}] &= - 
    2 M^2 \epsilon^{edfg} g_{c(a}^{(0)} g_{b)d}^{(0)} \dual R^{(0)}{}^{ch} {}_{fg}\nabla_h^{(0)}  \nabla^{(0)}_e \vartheta^{(1)} \\
    & \quad \quad  \quad  \quad + T_{ab}[\vartheta^{(1)}]\,.
\end{align}

Note the equations at order $\eps^k$ are homogenous in $\agb^k$, and hence~\cite{Witek:2018dmd} scales out the $\agb$ dependence. In order for all results to be physically meaningful, we must reintroduce $\agb$ (cf. for example~\cite{MashaHeadOn}). For the remainder of this paper, let us refer to $\Delta g_{ab}$ and $\Delta \vartheta$ as the leading-order corrections to the spacetime metric and the scalar field that we numerically evolve (with $M = 1$, $\agb$ scaled out).

\subsection{Initial data}

The background metric, $g_{ab}^{(0)}$, is given by an analytical Kerr solution in Kerr-Schild coordinates~\cite{MTW} (and is not evolved during the simulation). We solve for $\Delta \vartheta$ on the initial slice of this background solution using the methods provided in~\cite{Stein:2014xba} (note that the method of~\cite{Stein:2014xba} is not an expansion in spin, and is valid for all spins). We solve for $\Delta g_{ab}$, the leading order EDGB spacetime metric using the methods given in~\cite{MashaIDPaper}, which are valid for all GR background spacetimes. 

\subsection{Evolution}

We use the Spectral Einstein Code (SpEC) for our evolutions, a pseudo-spectral code that guarantees exponential numerical convergence in the fields~\cite{SpECwebsite}. We evolve $\Delta \vartheta$ using the method given in~\cite{Okounkova:2017yby}, a first-order, constraint-damping scheme valid for any background. We evolve $\Delta g_{ab}$ using the formalism outlined and tested in~\cite{MashaEvPaper}, a first-order, constraint-damping evolution system based on the generalized harmonic formalism~\cite{Lindblom2006}. This formalism was also used to perform head-on collisions of binary black holes in dCS~\cite{MashaHeadOn}. The domain consists of a series of spherical shells conforming to the shape of the horizon of the black hole, with the region inside of the horizon excised from the computational domain (the same setup as in~\cite{MashaEvPaper}). More details regarding the computational domain, spectral basis functions, filtering, and constraint damping parameters can be found in~\cite{Okounkova:2017yby, MashaEvPaper, MashaHeadOn}. 

\section{Results}

\begin{figure}
  \includegraphics[width=\columnwidth]{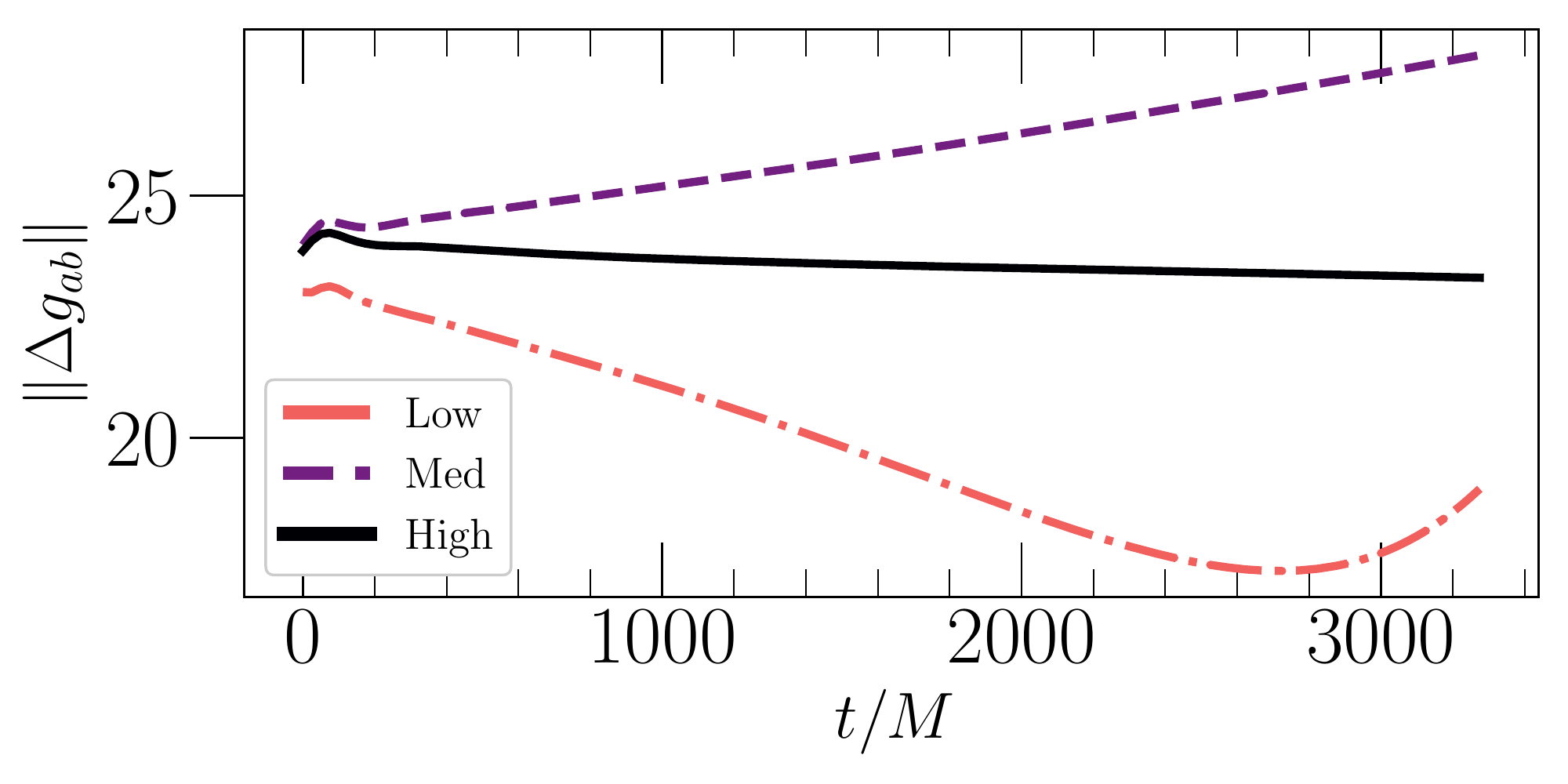}
  \caption{L2 norm of $\Delta g_{ab}$, the leading-order EDGB correction to the spacetime metric, as a function of simulation coordinate time. We evolve $\Delta g_{ab}$ on a stationary Kerr black hole background with dimensionless spin $\chi = 0.1$. Each curve corresponds to a different numerical resolution, where each successive resolution has two more spectral basis functions in the angular and radial directions in our spectral evolution code~\cite{SpECwebsite}. We see that while low numerical resolution exhibits linear growth is time, this growth goes away with increasing numerical resolution.  
  }
  \label{fig:dtg_0.1}
\end{figure}

In Fig.~\ref{fig:dtg_0.1}, we show the evolution of $\Delta g_{ab}$, the leading order EDGB correction to the spacetime metric as a function of time for a characteristic dimensionless spin of $\chi = 0.1$. We see that for low numerical resolutions, $\Delta g_{ab}$ increases linearly in time, suggesting a linear instability. However, this linear growth goes away as we increase the numerical resolution, thus showing that the black hole spacetime is numerically stable. We find similar results for the other spins simulated in this work, $\chi = 0, 0.6, 0.9$.

\begin{figure}
  \includegraphics[width=\columnwidth]{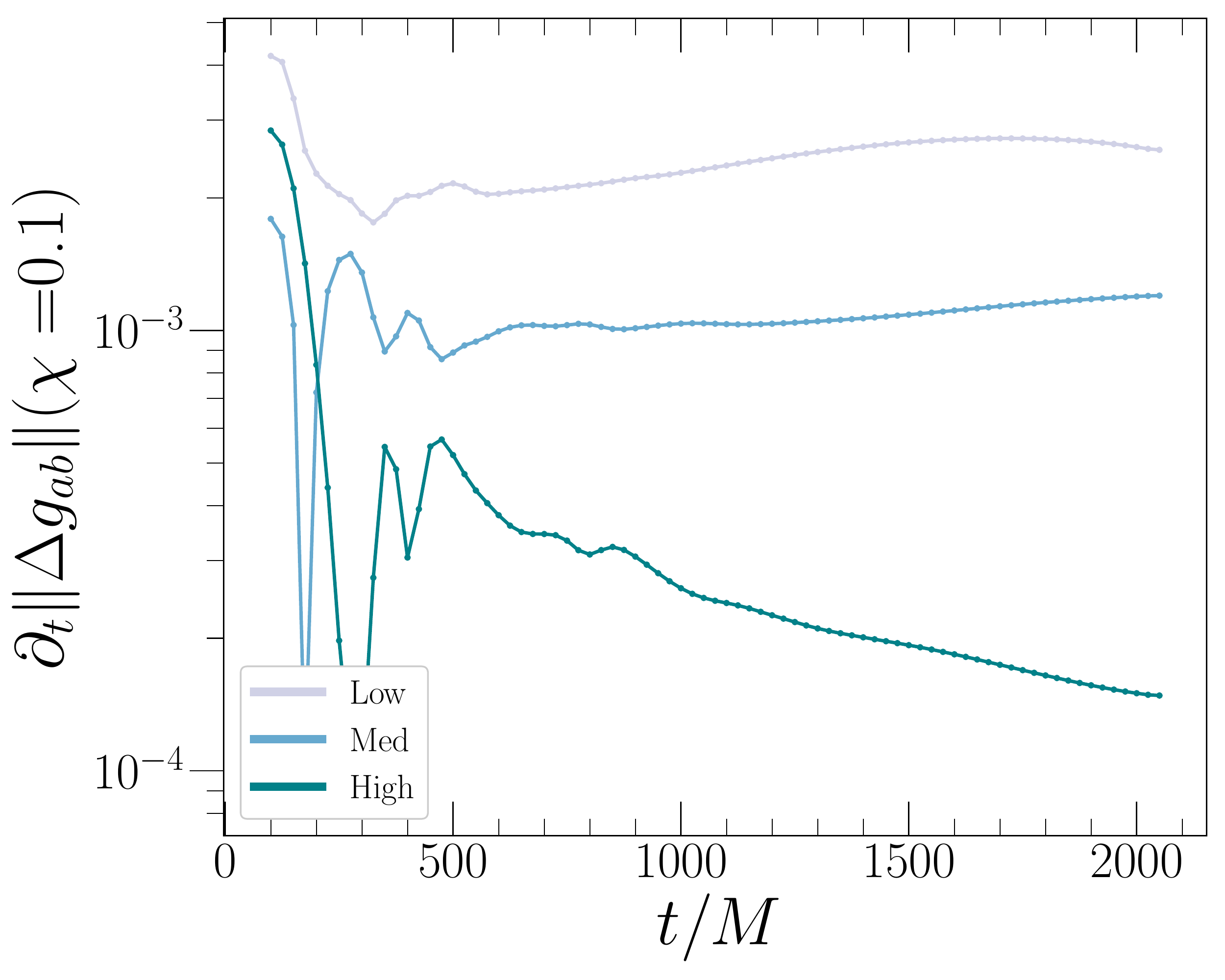}
  \caption{Time derivative of the norm of the EDGB metric deformation on a background with spin $\chi = 0.1$. We plot $\pd_t \| \Delta g_{ab}\|$, the time derivative of the curve in Fig.~\ref{fig:dtg_0.1}, for low, medium, and high resolution cases (where each successive resolution adds two radial and two spherical basis functions to our computation domain of spherical shells). We see that after an initial period of oscillation, the time derivative is convergent towards zero with increasing numerical resolution. 
  }
  \label{fig:Resolutions_0p1}
\end{figure}

\begin{figure}
  \includegraphics[width=\columnwidth]{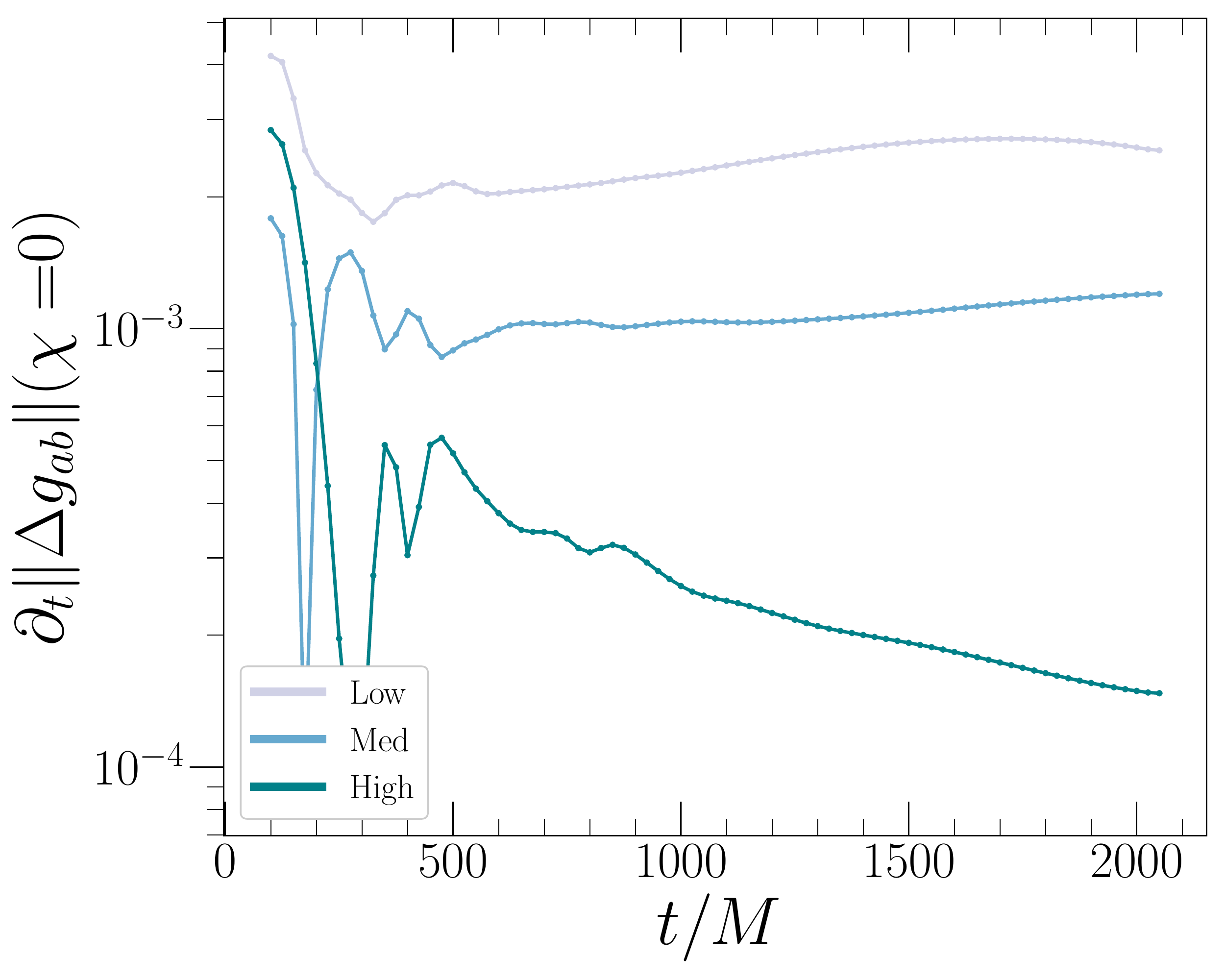}
  \caption{Same as Fig.~\ref{fig:Resolutions_0p1}, but for 
  a dimensionless background spin of $\chi = 0$. 
  }
  \label{fig:Resolutions_0}
\end{figure}

\begin{figure}
  \includegraphics[width=\columnwidth]{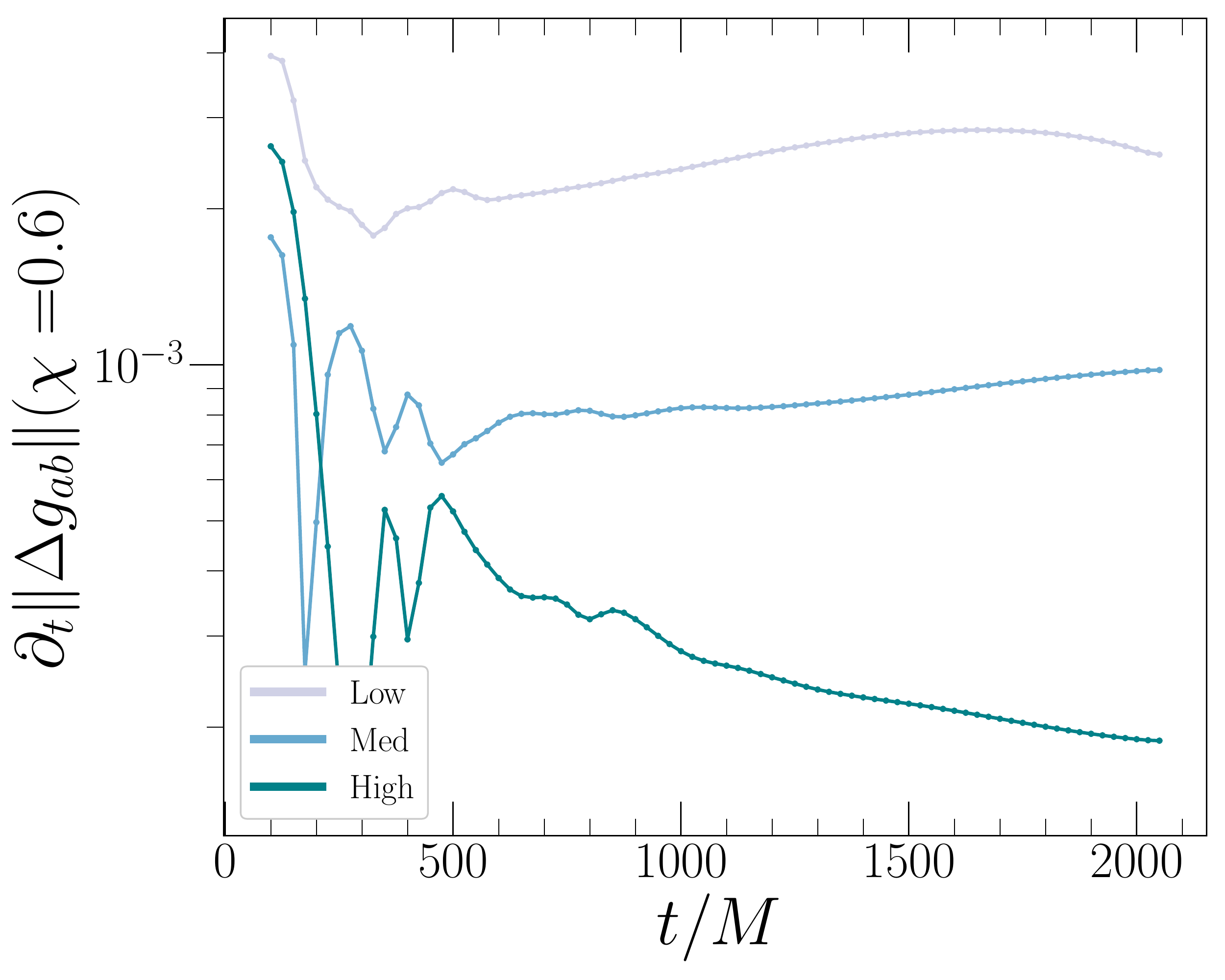}
  \caption{Same as Fig.~\ref{fig:Resolutions_0p1}, but for 
  a dimensionless background spin of $\chi = 0.6$. 
  }
  \label{fig:Resolutions_0p6}
\end{figure}

\begin{figure}
  \includegraphics[width=\columnwidth]{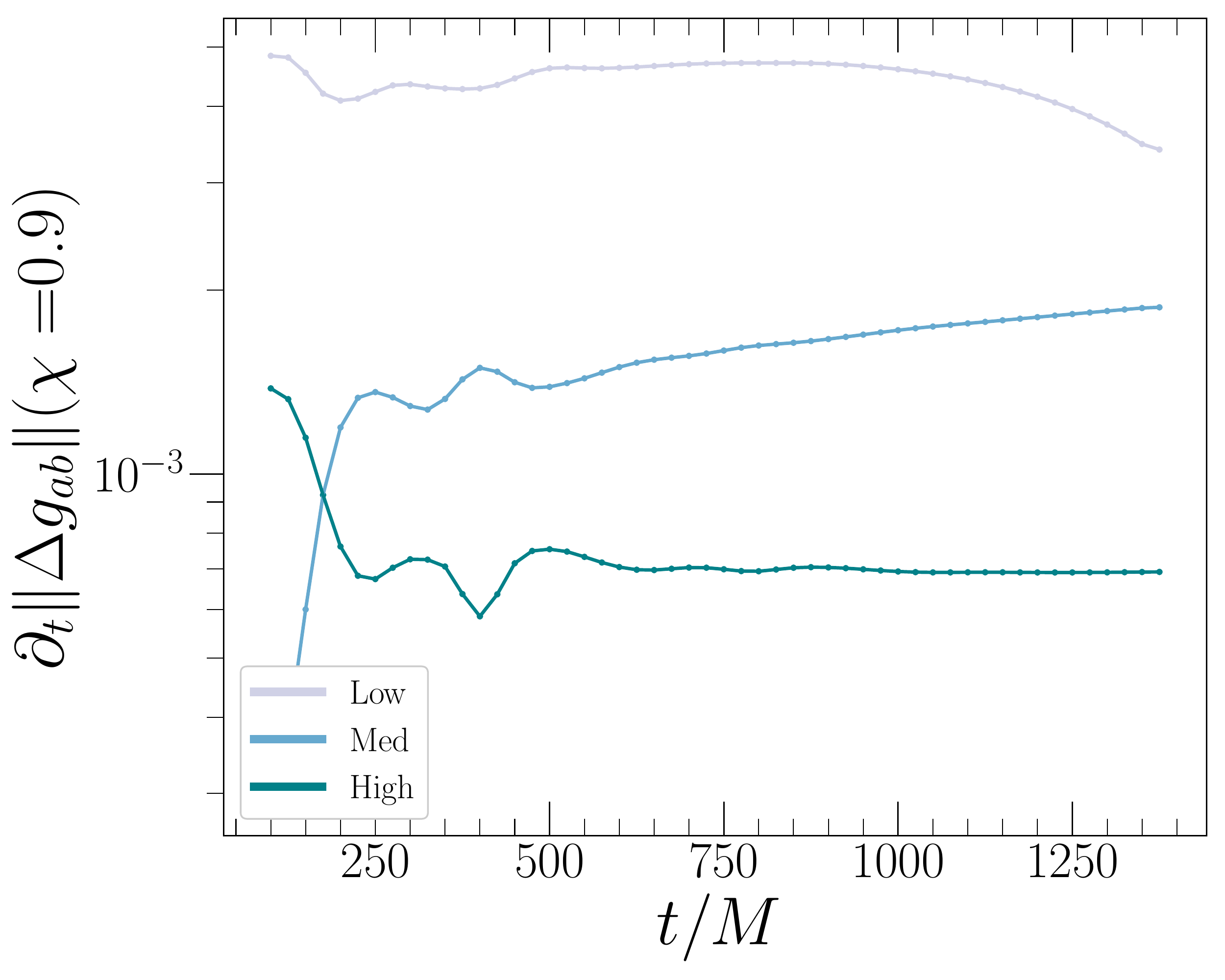}
  \caption{Same as Fig.~\ref{fig:Resolutions_0p1}, but for 
  a dimensionless background spin of $\chi = 0.9$. 
  }
  \label{fig:Resolutions_0p9}
\end{figure}

In Figs.~\ref{fig:Resolutions_0p1},~\ref{fig:Resolutions_0},~\ref{fig:Resolutions_0p6}, and~\ref{fig:Resolutions_0p9}, we take a closer look at the behavior of the EDGB metric deformation, showing the time derivative of $\Delta g_{ab}$. In each case, we see that as we increase numerical resolution, $\pd_t \Delta g_{ab}$ exponentially decreases towards zero. Thus, with increasing resolution, we can stably evolve rotating black holes in order-reduced EDGB. 

Note that the level of $\pd_t \Delta g_{ab}$ is greater for higher spins for the same numerical resolution, which is due to the fact that, in our coordinate system, black holes with higher spin require more spectral basis functions to be resolved with the same accuracy as black holes with lower spins~\cite{Lovelace:2010ne}. Similarly, higher-spin systems have lower (albeit still exponential) convergence rates than lower-spin systems~\cite{Lovelace:2010ne}. Given that we observe that the level of linear growth in the EDGB correction to the spacetime metric converges away with numerical resolution, we thus conclude that rotating black holes in EDGB are stable to leading (second) order. 

We have evolved the black holes for $\sim 4000\,M$. The length of a full (though inspiral, merger, ringdown) simulation with GW150914-like parameters is $\sim 3300\,M$~\cite{Lovelace:2016uwp}, and thus we have shown that, for high enough resolution, we can stably numerically evolve single black holes over the length of a binary black hole simulation.

\subsection{Regime of validity}

The results that we have presented for $g_{ab}^{(2)}$ have the EDGB coupling parameter $\agb$ scaled out. For the order reduction scheme to be valid, we require that $g_{ab}^{(2)} \lesssim C g_{ab}^{(0)}$, for some constant $C < 1$. This in turn becomes a constraint on $\agb$, of the form $\agb \lesssim \sqrt{C \frac{g_{ab}^{(0)}}{\Delta g_{ab}}}$. For a choice of $C = 0.1$, we find that $\agb \sim \mathcal{O}(0.1)$. 

\section{Conclusion}

In this study, we have investigated the stability of rotating black holes in Einstein dilaton Gauss-Bonnet gravity, showing that they are numerically stable to leading (second) order, for all spins. This in turn means that we satisfy a necessary condition for stably simulating binary black hole mergers in EDGB, as the black holes during the early inspiral (which can be modelled as a superposition of two black hole spacetimes~\cite{Lovelace:2008hd}) and the remnant black hole will not exhibit instabilities for high-enough numerical resolution. Moreover, we have evolved single black holes on timescales longer that a simulation with GW150914-like parameters~\cite{Lovelace:2016uwp}, showing that we can maintain this stability throughout a numerical simulation. 

In~\cite{MashaEvPaper}, we demonstrated the leading-order stability of rotating black holes in dCS gravity, and subsequently used the evolution scheme presented in~\cite{MashaEvPaper} to evolve the leading-order dCS metric deformation to a head-on binary black hole collision spacetime, computing the leading-order corrections to the quasi-normal mode spectrum~\cite{MashaHeadOn}. We can thus repeat~\cite{MashaHeadOn} for order-reduced EDGB gravity. 

The ultimate goal of this beyond-GR research program is to produce numerical relativity beyond-GR binary black hole waveforms through full inspiral, merger, and ringdown that are useful for model-dependent tests of general relativity. In this paper, by demonstrating stability of single, rotating black holes in EDGB, we have performed a necessary step in this program.

\section*{Acknowledgements}

We would like to thank Leo Stein for providing the code used to generate the scalar field initial data in~\cite{Stein:2014xba}. The Flatiron Institute is supported by the Simons Foundation. Computations were performed using the Spectral Einstein Code~\cite{SpECwebsite}. All computations were performed on the Wheeler cluster at Caltech, which is supported by the Sherman Fairchild Foundation and by Caltech.

\bibliography{biblio}

\begin{thebibliography}{26}%
\makeatletter
\providecommand \@ifxundefined [1]{%
 \@ifx{#1\undefined}
}%
\providecommand \@ifnum [1]{%
 \ifnum #1\expandafter \@firstoftwo
 \else \expandafter \@secondoftwo
 \fi
}%
\providecommand \@ifx [1]{%
 \ifx #1\expandafter \@firstoftwo
 \else \expandafter \@secondoftwo
 \fi
}%
\providecommand \natexlab [1]{#1}%
\providecommand \enquote  [1]{``#1''}%
\providecommand \bibnamefont  [1]{#1}%
\providecommand \bibfnamefont [1]{#1}%
\providecommand \citenamefont [1]{#1}%
\providecommand \href@noop [0]{\@secondoftwo}%
\providecommand \href [0]{\begingroup \@sanitize@url \@href}%
\providecommand \@href[1]{\@@startlink{#1}\@@href}%
\providecommand \@@href[1]{\endgroup#1\@@endlink}%
\providecommand \@sanitize@url [0]{\catcode `\\12\catcode `\$12\catcode
  `\&12\catcode `\#12\catcode `\^12\catcode `\_12\catcode `\%12\relax}%
\providecommand \@@startlink[1]{}%
\providecommand \@@endlink[0]{}%
\providecommand \url  [0]{\begingroup\@sanitize@url \@url }%
\providecommand \@url [1]{\endgroup\@href {#1}{\urlprefix }}%
\providecommand \urlprefix  [0]{URL }%
\providecommand \Eprint [0]{\href }%
\providecommand \doibase [0]{http://dx.doi.org/}%
\providecommand \selectlanguage [0]{\@gobble}%
\providecommand \bibinfo  [0]{\@secondoftwo}%
\providecommand \bibfield  [0]{\@secondoftwo}%
\providecommand \translation [1]{[#1]}%
\providecommand \BibitemOpen [0]{}%
\providecommand \bibitemStop [0]{}%
\providecommand \bibitemNoStop [0]{.\EOS\space}%
\providecommand \EOS [0]{\spacefactor3000\relax}%
\providecommand \BibitemShut  [1]{\csname bibitem#1\endcsname}%
\let\auto@bib@innerbib\@empty
\bibitem [{\citenamefont {Abbott}\ \emph {et~al.}(2016)\citenamefont {Abbott}
  \emph {et~al.}}]{TheLIGOScientific:2016src}%
  \BibitemOpen
  \bibfield  {author} {\bibinfo {author} {\bibfnamefont {B.~P.}\ \bibnamefont
  {Abbott}} \emph {et~al.} (\bibinfo {collaboration} {Virgo, LIGO
  Scientific}),\ }\href {\doibase 10.1103/PhysRevLett.116.221101,
  10.1103/PhysRevLett.121.129902} {\bibfield  {journal} {\bibinfo  {journal}
  {Phys. Rev. Lett.}\ }\textbf {\bibinfo {volume} {116}},\ \bibinfo {pages}
  {221101} (\bibinfo {year} {2016})},\ \bibinfo {note} {[Erratum: Phys. Rev.
  Lett.121,no.12,129902(2018)]},\ \Eprint {http://arxiv.org/abs/1602.03841}
  {arXiv:1602.03841 [gr-qc]} \BibitemShut {NoStop}%
\bibitem [{\citenamefont {Abbott}\ \emph {et~al.}(2019)\citenamefont {Abbott}
  \emph {et~al.}}]{LIGOScientific:2019fpa}%
  \BibitemOpen
  \bibfield  {author} {\bibinfo {author} {\bibfnamefont {B.~P.}\ \bibnamefont
  {Abbott}} \emph {et~al.} (\bibinfo {collaboration} {LIGO Scientific,
  Virgo}),\ }\href@noop {} {\  (\bibinfo {year} {2019})},\ \Eprint
  {http://arxiv.org/abs/1903.04467} {arXiv:1903.04467 [gr-qc]} \BibitemShut
  {NoStop}%
\bibitem [{\citenamefont {Healy}\ \emph {et~al.}(2012)\citenamefont {Healy},
  \citenamefont {Bode}, \citenamefont {Haas}, \citenamefont {Pazos},
  \citenamefont {Laguna}, \citenamefont {Shoemaker},\ and\ \citenamefont
  {Yunes}}]{Healy:2011ef}%
  \BibitemOpen
  \bibfield  {author} {\bibinfo {author} {\bibfnamefont {J.}~\bibnamefont
  {Healy}}, \bibinfo {author} {\bibfnamefont {T.}~\bibnamefont {Bode}},
  \bibinfo {author} {\bibfnamefont {R.}~\bibnamefont {Haas}}, \bibinfo {author}
  {\bibfnamefont {E.}~\bibnamefont {Pazos}}, \bibinfo {author} {\bibfnamefont
  {P.}~\bibnamefont {Laguna}}, \bibinfo {author} {\bibfnamefont {D.~M.}\
  \bibnamefont {Shoemaker}}, \ and\ \bibinfo {author} {\bibfnamefont
  {N.}~\bibnamefont {Yunes}},\ }\href {\doibase 10.1088/0264-9381/29/23/232002}
  {\bibfield  {journal} {\bibinfo  {journal} {Class. Quant. Grav.}\ }\textbf
  {\bibinfo {volume} {29}},\ \bibinfo {pages} {232002} (\bibinfo {year}
  {2012})},\ \Eprint {http://arxiv.org/abs/1112.3928} {arXiv:1112.3928 [gr-qc]}
  \BibitemShut {NoStop}%
\bibitem [{\citenamefont {Okounkova}\ \emph {et~al.}(2017)\citenamefont
  {Okounkova}, \citenamefont {Stein}, \citenamefont {Scheel},\ and\
  \citenamefont {Hemberger}}]{Okounkova:2017yby}%
  \BibitemOpen
  \bibfield  {author} {\bibinfo {author} {\bibfnamefont {M.}~\bibnamefont
  {Okounkova}}, \bibinfo {author} {\bibfnamefont {L.~C.}\ \bibnamefont
  {Stein}}, \bibinfo {author} {\bibfnamefont {M.~A.}\ \bibnamefont {Scheel}}, \
  and\ \bibinfo {author} {\bibfnamefont {D.~A.}\ \bibnamefont {Hemberger}},\
  }\href {\doibase 10.1103/PhysRevD.96.044020} {\bibfield  {journal} {\bibinfo
  {journal} {Phys. Rev.}\ }\textbf {\bibinfo {volume} {D96}},\ \bibinfo {pages}
  {044020} (\bibinfo {year} {2017})},\ \Eprint
  {http://arxiv.org/abs/1705.07924} {arXiv:1705.07924 [gr-qc]} \BibitemShut
  {NoStop}%
\bibitem [{\citenamefont {Witek}\ \emph {et~al.}(2018)\citenamefont {Witek},
  \citenamefont {Gualtieri}, \citenamefont {Pani},\ and\ \citenamefont
  {Sotiriou}}]{Witek:2018dmd}%
  \BibitemOpen
  \bibfield  {author} {\bibinfo {author} {\bibfnamefont {H.}~\bibnamefont
  {Witek}}, \bibinfo {author} {\bibfnamefont {L.}~\bibnamefont {Gualtieri}},
  \bibinfo {author} {\bibfnamefont {P.}~\bibnamefont {Pani}}, \ and\ \bibinfo
  {author} {\bibfnamefont {T.~P.}\ \bibnamefont {Sotiriou}},\ }\href@noop {} {\
   (\bibinfo {year} {2018})},\ \Eprint {http://arxiv.org/abs/1810.05177}
  {arXiv:1810.05177 [gr-qc]} \BibitemShut {NoStop}%
\bibitem [{\citenamefont {Okounkova}\ \emph
  {et~al.}(2019{\natexlab{a}})\citenamefont {Okounkova}, \citenamefont {Stein},
  \citenamefont {Scheel},\ and\ \citenamefont {Teukolsky}}]{MashaHeadOn}%
  \BibitemOpen
  \bibfield  {author} {\bibinfo {author} {\bibfnamefont {M.}~\bibnamefont
  {Okounkova}}, \bibinfo {author} {\bibfnamefont {L.~C.}\ \bibnamefont
  {Stein}}, \bibinfo {author} {\bibfnamefont {M.~A.}\ \bibnamefont {Scheel}}, \
  and\ \bibinfo {author} {\bibfnamefont {S.~A.}\ \bibnamefont {Teukolsky}},\
  }\href@noop {} {\  (\bibinfo {year} {2019}{\natexlab{a}})},\ \Eprint
  {http://arxiv.org/abs/1906.08789} {arXiv:1906.08789 [gr-qc]} \BibitemShut
  {NoStop}%
\bibitem [{\citenamefont {{Gross}}\ and\ \citenamefont
  {{Sloan}}(1987)}]{1987NuPhB.291...41G}%
  \BibitemOpen
  \bibfield  {author} {\bibinfo {author} {\bibfnamefont {D.~J.}\ \bibnamefont
  {{Gross}}}\ and\ \bibinfo {author} {\bibfnamefont {J.~H.}\ \bibnamefont
  {{Sloan}}},\ }\href {\doibase 10.1016/0550-3213(87)90465-2} {\bibfield
  {journal} {\bibinfo  {journal} {Nuclear Physics B}\ }\textbf {\bibinfo
  {volume} {291}},\ \bibinfo {pages} {41} (\bibinfo {year} {1987})}\BibitemShut
  {NoStop}%
\bibitem [{\citenamefont {Moura}\ and\ \citenamefont
  {Schiappa}(2007)}]{Moura:2006pz}%
  \BibitemOpen
  \bibfield  {author} {\bibinfo {author} {\bibfnamefont {F.}~\bibnamefont
  {Moura}}\ and\ \bibinfo {author} {\bibfnamefont {R.}~\bibnamefont
  {Schiappa}},\ }\href {\doibase 10.1088/0264-9381/24/2/006} {\bibfield
  {journal} {\bibinfo  {journal} {Class. Quant. Grav.}\ }\textbf {\bibinfo
  {volume} {24}},\ \bibinfo {pages} {361} (\bibinfo {year} {2007})},\ \Eprint
  {http://arxiv.org/abs/hep-th/0605001} {arXiv:hep-th/0605001 [hep-th]}
  \BibitemShut {NoStop}%
\bibitem [{\citenamefont {Berti}\ \emph {et~al.}(2015)\citenamefont {Berti}
  \emph {et~al.}}]{Berti:2015itd}%
  \BibitemOpen
  \bibfield  {author} {\bibinfo {author} {\bibfnamefont {E.}~\bibnamefont
  {Berti}} \emph {et~al.},\ }\href {\doibase 10.1088/0264-9381/32/24/243001}
  {\bibfield  {journal} {\bibinfo  {journal} {Class. Quant. Grav.}\ }\textbf
  {\bibinfo {volume} {32}},\ \bibinfo {pages} {243001} (\bibinfo {year}
  {2015})},\ \Eprint {http://arxiv.org/abs/1501.07274} {arXiv:1501.07274
  [gr-qc]} \BibitemShut {NoStop}%
\bibitem [{\citenamefont {Delsate}\ \emph {et~al.}(2015)\citenamefont
  {Delsate}, \citenamefont {Hilditch},\ and\ \citenamefont
  {Witek}}]{Delsate:2014hba}%
  \BibitemOpen
  \bibfield  {author} {\bibinfo {author} {\bibfnamefont {T.}~\bibnamefont
  {Delsate}}, \bibinfo {author} {\bibfnamefont {D.}~\bibnamefont {Hilditch}}, \
  and\ \bibinfo {author} {\bibfnamefont {H.}~\bibnamefont {Witek}},\ }\href
  {\doibase 10.1103/PhysRevD.91.024027} {\bibfield  {journal} {\bibinfo
  {journal} {Phys. Rev.}\ }\textbf {\bibinfo {volume} {D91}},\ \bibinfo {pages}
  {024027} (\bibinfo {year} {2015})},\ \Eprint {http://arxiv.org/abs/1407.6727}
  {arXiv:1407.6727 [gr-qc]} \BibitemShut {NoStop}%
\bibitem [{\citenamefont {Kanti}\ \emph {et~al.}(1996)\citenamefont {Kanti},
  \citenamefont {Mavromatos}, \citenamefont {Rizos}, \citenamefont {Tamvakis},\
  and\ \citenamefont {Winstanley}}]{Kanti:1995vq}%
  \BibitemOpen
  \bibfield  {author} {\bibinfo {author} {\bibfnamefont {P.}~\bibnamefont
  {Kanti}}, \bibinfo {author} {\bibfnamefont {N.~E.}\ \bibnamefont
  {Mavromatos}}, \bibinfo {author} {\bibfnamefont {J.}~\bibnamefont {Rizos}},
  \bibinfo {author} {\bibfnamefont {K.}~\bibnamefont {Tamvakis}}, \ and\
  \bibinfo {author} {\bibfnamefont {E.}~\bibnamefont {Winstanley}},\ }\href
  {\doibase 10.1103/PhysRevD.54.5049} {\bibfield  {journal} {\bibinfo
  {journal} {Phys. Rev.}\ }\textbf {\bibinfo {volume} {D54}},\ \bibinfo {pages}
  {5049} (\bibinfo {year} {1996})},\ \Eprint
  {http://arxiv.org/abs/hep-th/9511071} {arXiv:hep-th/9511071 [hep-th]}
  \BibitemShut {NoStop}%
\bibitem [{\citenamefont {Pani}\ and\ \citenamefont
  {Cardoso}(2009)}]{Pani:2009wy}%
  \BibitemOpen
  \bibfield  {author} {\bibinfo {author} {\bibfnamefont {P.}~\bibnamefont
  {Pani}}\ and\ \bibinfo {author} {\bibfnamefont {V.}~\bibnamefont {Cardoso}},\
  }\href {\doibase 10.1103/PhysRevD.79.084031} {\bibfield  {journal} {\bibinfo
  {journal} {Phys. Rev.}\ }\textbf {\bibinfo {volume} {D79}},\ \bibinfo {pages}
  {084031} (\bibinfo {year} {2009})},\ \Eprint {http://arxiv.org/abs/0902.1569}
  {arXiv:0902.1569 [gr-qc]} \BibitemShut {NoStop}%
\bibitem [{\citenamefont {Ayzenberg}\ and\ \citenamefont
  {Yunes}(2014)}]{Ayzenberg:2014aka}%
  \BibitemOpen
  \bibfield  {author} {\bibinfo {author} {\bibfnamefont {D.}~\bibnamefont
  {Ayzenberg}}\ and\ \bibinfo {author} {\bibfnamefont {N.}~\bibnamefont
  {Yunes}},\ }\href {\doibase 10.1103/PhysRevD.91.069905,
  10.1103/PhysRevD.90.044066} {\bibfield  {journal} {\bibinfo  {journal} {Phys.
  Rev.}\ }\textbf {\bibinfo {volume} {D90}},\ \bibinfo {pages} {044066}
  (\bibinfo {year} {2014})},\ \bibinfo {note} {[Erratum: Phys.
  Rev.D91,no.6,069905(2015)]},\ \Eprint {http://arxiv.org/abs/1405.2133}
  {arXiv:1405.2133 [gr-qc]} \BibitemShut {NoStop}%
\bibitem [{\citenamefont {Kleihaus}\ \emph {et~al.}(2011)\citenamefont
  {Kleihaus}, \citenamefont {Kunz},\ and\ \citenamefont
  {Radu}}]{Kleihaus:2011tg}%
  \BibitemOpen
  \bibfield  {author} {\bibinfo {author} {\bibfnamefont {B.}~\bibnamefont
  {Kleihaus}}, \bibinfo {author} {\bibfnamefont {J.}~\bibnamefont {Kunz}}, \
  and\ \bibinfo {author} {\bibfnamefont {E.}~\bibnamefont {Radu}},\ }\href
  {\doibase 10.1103/PhysRevLett.106.151104} {\bibfield  {journal} {\bibinfo
  {journal} {Phys. Rev. Lett.}\ }\textbf {\bibinfo {volume} {106}},\ \bibinfo
  {pages} {151104} (\bibinfo {year} {2011})},\ \Eprint
  {http://arxiv.org/abs/1101.2868} {arXiv:1101.2868 [gr-qc]} \BibitemShut
  {NoStop}%
\bibitem [{\citenamefont {Kleihaus}\ \emph {et~al.}(2016)\citenamefont
  {Kleihaus}, \citenamefont {Kunz}, \citenamefont {Mojica},\ and\ \citenamefont
  {Radu}}]{Kleihaus:2015aje}%
  \BibitemOpen
  \bibfield  {author} {\bibinfo {author} {\bibfnamefont {B.}~\bibnamefont
  {Kleihaus}}, \bibinfo {author} {\bibfnamefont {J.}~\bibnamefont {Kunz}},
  \bibinfo {author} {\bibfnamefont {S.}~\bibnamefont {Mojica}}, \ and\ \bibinfo
  {author} {\bibfnamefont {E.}~\bibnamefont {Radu}},\ }\href {\doibase
  10.1103/PhysRevD.93.044047} {\bibfield  {journal} {\bibinfo  {journal} {Phys.
  Rev.}\ }\textbf {\bibinfo {volume} {D93}},\ \bibinfo {pages} {044047}
  (\bibinfo {year} {2016})},\ \Eprint {http://arxiv.org/abs/1511.05513}
  {arXiv:1511.05513 [gr-qc]} \BibitemShut {NoStop}%
\bibitem [{\citenamefont {Cunha}\ \emph {et~al.}(2017)\citenamefont {Cunha},
  \citenamefont {Herdeiro}, \citenamefont {Kleihaus}, \citenamefont {Kunz},\
  and\ \citenamefont {Radu}}]{Cunha:2016wzk}%
  \BibitemOpen
  \bibfield  {author} {\bibinfo {author} {\bibfnamefont {P.~V.~P.}\
  \bibnamefont {Cunha}}, \bibinfo {author} {\bibfnamefont {C.~A.~R.}\
  \bibnamefont {Herdeiro}}, \bibinfo {author} {\bibfnamefont {B.}~\bibnamefont
  {Kleihaus}}, \bibinfo {author} {\bibfnamefont {J.}~\bibnamefont {Kunz}}, \
  and\ \bibinfo {author} {\bibfnamefont {E.}~\bibnamefont {Radu}},\ }\href
  {\doibase 10.1016/j.physletb.2017.03.020} {\bibfield  {journal} {\bibinfo
  {journal} {Phys. Lett.}\ }\textbf {\bibinfo {volume} {B768}},\ \bibinfo
  {pages} {373} (\bibinfo {year} {2017})},\ \Eprint
  {http://arxiv.org/abs/1701.00079} {arXiv:1701.00079 [gr-qc]} \BibitemShut
  {NoStop}%
\bibitem [{\citenamefont {Kanti}\ \emph {et~al.}(1998)\citenamefont {Kanti},
  \citenamefont {Mavromatos}, \citenamefont {Rizos}, \citenamefont {Tamvakis},\
  and\ \citenamefont {Winstanley}}]{Kanti:1997br}%
  \BibitemOpen
  \bibfield  {author} {\bibinfo {author} {\bibfnamefont {P.}~\bibnamefont
  {Kanti}}, \bibinfo {author} {\bibfnamefont {N.~E.}\ \bibnamefont
  {Mavromatos}}, \bibinfo {author} {\bibfnamefont {J.}~\bibnamefont {Rizos}},
  \bibinfo {author} {\bibfnamefont {K.}~\bibnamefont {Tamvakis}}, \ and\
  \bibinfo {author} {\bibfnamefont {E.}~\bibnamefont {Winstanley}},\ }\href
  {\doibase 10.1103/PhysRevD.57.6255} {\bibfield  {journal} {\bibinfo
  {journal} {Phys. Rev.}\ }\textbf {\bibinfo {volume} {D57}},\ \bibinfo {pages}
  {6255} (\bibinfo {year} {1998})},\ \Eprint
  {http://arxiv.org/abs/hep-th/9703192} {arXiv:hep-th/9703192 [hep-th]}
  \BibitemShut {NoStop}%
\bibitem [{\citenamefont {Okounkova}\ \emph
  {et~al.}(2019{\natexlab{b}})\citenamefont {Okounkova}, \citenamefont
  {Scheel},\ and\ \citenamefont {Teukolsky}}]{MashaEvPaper}%
  \BibitemOpen
  \bibfield  {author} {\bibinfo {author} {\bibfnamefont {M.}~\bibnamefont
  {Okounkova}}, \bibinfo {author} {\bibfnamefont {M.~A.}\ \bibnamefont
  {Scheel}}, \ and\ \bibinfo {author} {\bibfnamefont {S.~A.}\ \bibnamefont
  {Teukolsky}},\ }\href {\doibase 10.1103/PhysRevD.99.044019} {\bibfield
  {journal} {\bibinfo  {journal} {Phys. Rev.}\ }\textbf {\bibinfo {volume}
  {D99}},\ \bibinfo {pages} {044019} (\bibinfo {year} {2019}{\natexlab{b}})},\
  \Eprint {http://arxiv.org/abs/1811.10713} {arXiv:1811.10713 [gr-qc]}
  \BibitemShut {NoStop}%
\bibitem [{\citenamefont {Misner}\ \emph {et~al.}(1973)\citenamefont {Misner},
  \citenamefont {Thorne},\ and\ \citenamefont {Wheeler}}]{MTW}%
  \BibitemOpen
  \bibfield  {author} {\bibinfo {author} {\bibfnamefont {C.~W.}\ \bibnamefont
  {Misner}}, \bibinfo {author} {\bibfnamefont {K.~S.}\ \bibnamefont {Thorne}},
  \ and\ \bibinfo {author} {\bibfnamefont {J.~A.}\ \bibnamefont {Wheeler}},\
  }\href@noop {} {\emph {\bibinfo {title} {Gravitation}}}\ (\bibinfo
  {publisher} {Freeman},\ \bibinfo {address} {New York, New York},\ \bibinfo
  {year} {1973})\BibitemShut {NoStop}%
\bibitem [{\citenamefont {Stein}(2014)}]{Stein:2014xba}%
  \BibitemOpen
  \bibfield  {author} {\bibinfo {author} {\bibfnamefont {L.~C.}\ \bibnamefont
  {Stein}},\ }\href {\doibase 10.1103/PhysRevD.90.044061} {\bibfield  {journal}
  {\bibinfo  {journal} {Phys. Rev.}\ }\textbf {\bibinfo {volume} {D90}},\
  \bibinfo {pages} {044061} (\bibinfo {year} {2014})},\ \Eprint
  {http://arxiv.org/abs/1407.2350} {arXiv:1407.2350 [gr-qc]} \BibitemShut
  {NoStop}%
\bibitem [{\citenamefont {Okounkova}\ \emph {et~al.}(2018)\citenamefont
  {Okounkova}, \citenamefont {Scheel},\ and\ \citenamefont
  {Teukolsky}}]{MashaIDPaper}%
  \BibitemOpen
  \bibfield  {author} {\bibinfo {author} {\bibfnamefont {M.}~\bibnamefont
  {Okounkova}}, \bibinfo {author} {\bibfnamefont {M.~A.}\ \bibnamefont
  {Scheel}}, \ and\ \bibinfo {author} {\bibfnamefont {S.~A.}\ \bibnamefont
  {Teukolsky}},\ }\href {\doibase 10.1088/1361-6382/aafcdf} {\  (\bibinfo
  {year} {2018}),\ 10.1088/1361-6382/aafcdf},\ \Eprint
  {http://arxiv.org/abs/1810.05306} {arXiv:1810.05306 [gr-qc]} \BibitemShut
  {NoStop}%
\bibitem [{SpE()}]{SpECwebsite}%
  \BibitemOpen
  \href@noop {} {\enquote {\bibinfo {title} {The {S}pectral {E}instein {C}ode
  ({SpEC})},}\ }\bibinfo {howpublished}
  {\url{http://www.black-holes.org/SpEC.html}}\BibitemShut {NoStop}%
\bibitem [{\citenamefont {Lindblom}\ \emph {et~al.}(2006)\citenamefont
  {Lindblom}, \citenamefont {Scheel}, \citenamefont {Kidder}, \citenamefont
  {Owen},\ and\ \citenamefont {Rinne}}]{Lindblom2006}%
  \BibitemOpen
  \bibfield  {author} {\bibinfo {author} {\bibfnamefont {L.}~\bibnamefont
  {Lindblom}}, \bibinfo {author} {\bibfnamefont {M.~A.}\ \bibnamefont
  {Scheel}}, \bibinfo {author} {\bibfnamefont {L.~E.}\ \bibnamefont {Kidder}},
  \bibinfo {author} {\bibfnamefont {R.}~\bibnamefont {Owen}}, \ and\ \bibinfo
  {author} {\bibfnamefont {O.}~\bibnamefont {Rinne}},\ }\href {\doibase
  10.1088/0264-9381/23/16/S09} {\bibfield  {journal} {\bibinfo  {journal}
  {Class. Quant. Grav.}\ }\textbf {\bibinfo {volume} {23}},\ \bibinfo {pages}
  {S447} (\bibinfo {year} {2006})},\ \Eprint
  {http://arxiv.org/abs/gr-qc/0512093} {arXiv:gr-qc/0512093 [gr-qc]}
  \BibitemShut {NoStop}%
\bibitem [{\citenamefont {Lovelace}\ \emph {et~al.}(2011)\citenamefont
  {Lovelace}, \citenamefont {Scheel},\ and\ \citenamefont
  {Szilagyi}}]{Lovelace:2010ne}%
  \BibitemOpen
  \bibfield  {author} {\bibinfo {author} {\bibfnamefont {G.}~\bibnamefont
  {Lovelace}}, \bibinfo {author} {\bibfnamefont {M.}~\bibnamefont {Scheel}}, \
  and\ \bibinfo {author} {\bibfnamefont {B.}~\bibnamefont {Szilagyi}},\ }\href
  {\doibase 10.1103/PhysRevD.83.024010} {\bibfield  {journal} {\bibinfo
  {journal} {Phys. Rev.}\ }\textbf {\bibinfo {volume} {D83}},\ \bibinfo {pages}
  {024010} (\bibinfo {year} {2011})},\ \Eprint {http://arxiv.org/abs/1010.2777}
  {arXiv:1010.2777 [gr-qc]} \BibitemShut {NoStop}%
\bibitem [{\citenamefont {Lovelace}\ \emph {et~al.}(2016)\citenamefont
  {Lovelace} \emph {et~al.}}]{Lovelace:2016uwp}%
  \BibitemOpen
  \bibfield  {author} {\bibinfo {author} {\bibfnamefont {G.}~\bibnamefont
  {Lovelace}} \emph {et~al.},\ }\href {\doibase 10.1088/0264-9381/33/24/244002}
  {\bibfield  {journal} {\bibinfo  {journal} {Class. Quant. Grav.}\ }\textbf
  {\bibinfo {volume} {33}},\ \bibinfo {pages} {244002} (\bibinfo {year}
  {2016})},\ \Eprint {http://arxiv.org/abs/1607.05377} {arXiv:1607.05377
  [gr-qc]} \BibitemShut {NoStop}%
\bibitem [{\citenamefont {Lovelace}(2009)}]{Lovelace:2008hd}%
  \BibitemOpen
  \bibfield  {author} {\bibinfo {author} {\bibfnamefont {G.}~\bibnamefont
  {Lovelace}},\ }\bibfield  {booktitle} {\emph {\bibinfo {booktitle}
  {{Numerical relativity data analysis. Proceedings, 2nd Meeting, NRDA 2008,
  Syracuse, USA, August 11-14, 2008}}},\ }\href {\doibase
  10.1088/0264-9381/26/11/114002} {\bibfield  {journal} {\bibinfo  {journal}
  {Class. Quant. Grav.}\ }\textbf {\bibinfo {volume} {26}},\ \bibinfo {pages}
  {114002} (\bibinfo {year} {2009})},\ \Eprint {http://arxiv.org/abs/0812.3132}
  {arXiv:0812.3132 [gr-qc]} \BibitemShut {NoStop}%
\end{thebibliography}%
\end{document}